\documentclass[a4paper,conference]{IEEEtran}
\IEEEoverridecommandlockouts
\usepackage{cite}
\usepackage{multirow}
\usepackage{colortbl}
\usepackage{hhline}
\usepackage{amsmath,amssymb,amsfonts}
\usepackage{graphicx} 
\usepackage{caption,subcaption}
\usepackage{algorithmic}
\usepackage{graphicx}
\usepackage{textcomp}
\usepackage{xcolor}
\usepackage{tikz}
\usepackage{siunitx}
\usepackage{pdfpages}

\usepackage[protrusion=true]{microtype}
\usepackage{setspace}
\usetikzlibrary{shapes.geometric, positioning, calc, decorations.pathreplacing, shapes.misc, arrows.meta, decorations.markings}
\usepackage[RPvoltages]{circuitikz}
\tikzset{font=\fontsize{52}{58}}

\newcommand\verticaloffsetxbar{2cm}
\newcommand\horizontaloffsetxbar{2cm }
\newcommand\scaleresistances{0.9} 
\newcommand\distancePEbox{2.5cm }

\newcommand\widthxbarbox{2*\horizontaloffsetxbar +0.8 cm}
\newcommand\heightxbarbox{2*\verticaloffsetxbar}
\newcommand\widthxbarredbox{2* \horizontaloffsetxbar+ 0.3 cm }
\newcommand\heightxbarredbox{2*\verticaloffsetxbar}
\newcommand\widthxbarpurplebox{1*\horizontaloffsetxbar + 0.3 cm  }
\newcommand\heightxbarpurplebox{\heightxbarredbox}
\newcommand\widthPEbox{(\widthxbarbox+\widthxbarredbox)*1.2}
\newcommand\heightPEbox{\heightxbarredbox*2.36}

\def\BibTeX{{\rm B\kern-.05em{\sc i\kern-.025em b}\kern-.08em
    T\kern-.1667em\lower.7ex\hbox{E}\kern-.125emX}}

\usepackage[all]{background}
\usepackage{stackengine}
\setstackEOL{\\}
\setstackgap{L}{\normalbaselineskip}
\SetBgContents{\color{gray}{\tiny \Longstack{\copyright$ $ 2024 IEEE. PREPRINT - To appear at 29th IEEE European Test Symposium 2024 (ETS) 2024.}}}
\SetBgPosition{4.5cm,1cm}
\SetBgOpacity{1.0}
\SetBgAngle{0}
\SetBgScale{1.8}

\begin{document}
\bstctlcite{MyBSTcontrol}

\title{Error Detection and Correction Codes for Safe In-Memory Computations\\
}

\author{\IEEEauthorblockN{Luca~Parrini\IEEEauthorrefmark{1}\IEEEauthorrefmark{4}, Taha~Soliman\IEEEauthorrefmark{1}, Benjamin~Hettwer\IEEEauthorrefmark{1}, Jan~Micha~Borrmann\IEEEauthorrefmark{1}, Simranjeet~Singh\IEEEauthorrefmark{2},\\ Ankit~Bende\IEEEauthorrefmark{2}, Vikas~Rana\IEEEauthorrefmark{2}, Farhad~Merchant\IEEEauthorrefmark{3}, Norbert~Wehn\IEEEauthorrefmark{4}}
    \IEEEauthorblockA{\IEEEauthorrefmark{1}Bosch Corporate Research, Robert Bosch GmbH, Germany \IEEEauthorrefmark{2}Forschungszentrum Jülich GmbH, Germany \\ \IEEEauthorrefmark{3}Newcastle University, UK  \IEEEauthorrefmark{4}RPTU Kaiserslautern-Landau, Germany\\Email: luca.parrini@de.bosch.com}
}

\maketitle

\begin{abstract}
In-Memory Computing (IMC) introduces a new paradigm of computation that offers high efficiency in terms of latency and power consumption for AI accelerators. However, the non-idealities and defects of emerging technologies used in advanced IMC can severely degrade the accuracy of inferred Neural Networks (NN) and lead to malfunctions in safety-critical applications. In this paper, we investigate an architectural-level mitigation technique based on the coordinated action of multiple checksum codes, to detect and correct errors at run-time.
This implementation demonstrates higher efficiency in recovering accuracy across different AI algorithms and technologies compared to more traditional methods such as Triple Modular Redundancy (TMR).    
The results show that several configurations of our implementation recover more than 91\% of the original accuracy with less than half of the area required by TMR and less than 40\% of latency overhead.
\end{abstract}


\section{Introduction}
In-Memory Computing (IMC) based AI accelerators greatly reduce the latency and power consumption of an embedded chip, as they can overcome the bottleneck caused by the data transfers from the memory systems to the computation units. However, several factors can influence the resistive values of the emerging devices that represent the state of the non-volatile memory cells in these architectures \cite{yan_reliability_2023}. The accuracy degradation resulting from the propagation of these fluctuations into system failures is considered unacceptable for safety-critical applications. State-of-the-art techniques such as error detection/correction codes that only necessitate modifications to the hardware architecture, have the advantage of being easily scalable and technology-agnostic. Prior studies in this direction have introduced two distinct types of codes re-adapted for IMC, namely non-arithmetic and arithmetic. The latter are more suitable in the context of AI, as they aim to correct errors that appear as additive syndromes in the output without imposing significant constraints on the dimensions and technology of the crossbar.
However, solutions such as AN codes \cite{feinberg_making_2018} and checksums \cite{das_selective_2020, hemaram_adaptive_2022, liu_fault_2018}, do not fully exploit the inherent fault tolerance of Neural Network (NN) algorithms. As a result, the implementation often requires an unnecessary amount of redundant logic.

This paper presents a "\textit{Neural Checksums}" implementation designed specifically for the trending IMC technology, where the level of integrity protection can be tuned to target only specific neurons or bits of the running NN. 
In contrast to previous works on correction techniques for IMC, the methodology presented in this paper is accuracy-driven. Our error correction routine is tailored to prevent the propagation of errors to the final accuracy of the NN architecture.
This approach allows us to achieve an accuracy close to the baseline of the inferred algorithm. The proposed checksum method consistently outperforms more traditional methods, such as Triple Modular Redundancy (TMR), while incurring a comparable area overhead to other state-of-the-art solutions.
Moreover, in comparison to these techniques fine-tuned to specific fault distributions, our contribution is error-agnostic. In particular, our new contribution can be summarized as follows:
\begin{itemize}
    \item Two distinct checksum blocks in the form of IMC crossbar are introduced to perform real-time detection and correction of any arithmetic errors. 
    \item The safety blocks function independently to ensure the correct mitigation of errors and minimize the risk of malfunctions.
    \item Our methodology allows trading-off area and latency overhead for accuracy based on the requirements of the target application.
\end{itemize}

\section{Background}
\label{sec:Background}
Convolutional Neural Networks are a class of DNNs employed for a wide range of applications including image processing \cite{soliman_felix_2022}.  
The calculation of each value $O(x,y,z)$ of the output feature map is determined by the following equation:
\begin{equation}
\label{eq:nn_comp}
    O(x,y,z) = \sum_{l=0}^{s-1}\sum_{j=0}^{s-1}\sum_{h=0}^{In_d}In(x+l, y+j, h)w_z(l,j,h) 
\end{equation}
Where the parameters $In$ and $w_z$ represent the inputs and the weights of the kernel at depth $z$, respectively. The indexes $s$ and $In_d$ instead represent the kernel size and the depth of the input feature map, respectively.
\subsection{IMC-based AI Accelerators and Variations}
State-of-the-art IMC-based embedded AI accelerators distribute the workload described by Equation \ref{eq:nn_comp} into a matrix-like structure of distinct computational units called "Processing Elements" (PEs).
These units encapsulate a grid of memory cells, each of which maps a partial bit representation of the pre-trained weights as resistive values. As these cells share a common vertical and horizontal output wire in a crossbar-like structure, they can be used to perform Multiply and Accumulate (MAC) arithmetic operations in various domains.
The analog/mixed domains allow for a higher degree of parallelism, suitable for AI use cases. However, the computations are also the most heavily impacted by the variations of the resistive values of the cells. These defects may result in stuck-at-faults in their state or transient errors in the outputs. While the former can be easily mitigated with any fault-aware mapping scheme, the latter are much harder to predict due to their random nature, such as the Telegraph and Environment noise \cite{yan_reliability_2023}. This paper evaluates the fault tolerance of a safety mechanism that combines checksum codes and time redundancy, using two emerging technologies used today to implement resistive crossbars, Ferroelectric Field-Effect Transistor (FeFET) and Resistive Random Access Memory (RRAM).
\subsubsection{FeFET}
FeFET has emerged as a robust option for implementing non-volatile cells due to its advantageous properties in terms of required area, power consumption, and read/write latency \cite{torwards_taha}. They maintain a similar behavior to one of the typical MOSFET cells, except for the added $HfO_2$ substrate at his gate, which allows the electrical programmability of the $V_{th}$ of the device. Once programmed, the value of the $I_{ds}$ current of the cell behaves similarly to the 'AND' boolean gate when a binary signal is applied. This mechanism was exploited in several architectures, including FELIX \cite{soliman_felix_2022}, over which we simulate the variations for cells having dimensions of $W\times L$ = $500\times 500$ $nm^2$ \cite{8510622}. We applied the same error model as \cite{torwards_taha} based on random soft faults, which also proposes a mitigation technique limited to architectures with this type of technology. 

\subsubsection{RRAM}
Promising for their high-density storage, low power consumption, and accelerated read/write speeds, RRAM technologies show great potential in advancing memory capabilities~\cite{waserReram}. The RRAM memristive device utilized in this study comprises a Pt/TaO$\rm _x$/W/Pt stack, with dimensions of $100\times 100$ $nm^2$~\cite{bende2023experimental}. These devices are seamlessly integrated into CMOS 180 $nm$ technology provided by X-FAB, forming active memristive cells known as 1T1R cells. The resistive value of these cells is also programmable to a high (HRS) or low (LRS) resistance state. Each device undergoes 100 cycles for variability assessment on a RRAM crossbar, and the simulation was carried out using the error model designed according to the architecture outlined in~\cite{Bengel_2022}.

\section{Checksums Methodology}
\label{sec:methodology}
Previous studies on the application of checksums to AI, such as \cite{ozen_sanity-check_2019}, exploit the linearity of the MAC computations within each neuron to perform the redundant operations. This paper extends this method to create two separate safety blocks consisting of the checksum codes shown in Fig. \ref{fig:myarch}. By exploiting the inputs’ reuse between parameters on different levels of abstraction, they perform the computations independently and work together to detect and correct errors in the design at runtime.
Our method is explained in the following equation:
\begin{equation}
\begin{gathered}
\label{eq:proof_occadd}
    O^{acc} = \sum_{i}O_{i} = \sum_{i}\sum_{k}In_{k} \cdot w_{i,k} = \\ \sum_{k}In_{k}\sum_{i}w_{i,k} =  \sum_{k}In_{k} \cdot W_k^{ch} = O_i^{checksum} 
\end{gathered}
\end{equation}
$O^{acc}$ corresponds to the sum over a generic index $i$ of the partial results $O_{i}$ calculated by each column of the crossbar. $W_k^{ch}$ represents the summation of the weight values over the same index $i$, mapped onto the safety block, which then calculates the checksum results $O_i^{checksum}$ accordingly. The index $k$ corresponds to the rows' number of each crossbar while $i$ can be either the index of each column inside each crossbar ($b$) or the one of the respective PE ($n$) as shown in Fig. \ref{fig:myarch}.  
It is possible to distinguish these checksum codes according to the level of abstraction at which they have been implemented. 
\subsection{Crossbar Checksum}
This code is called "crossbar checksum" ($O_n^{Crossch}$) as it is implemented inside each PE block in the form of redundant columns. These additional cells determine the values that correspond to the sum of the parameters for each row in a specific section of the crossbar ($W^{Crossch}$), as illustrated in Fig. \ref{fig:myarch}. Each time a computation is completed and the output of the MAC operation from each column of the crossbar inside a PE is finalized, the values are summed together in the digital domain using an adder tree. In the meanwhile, since the redundant columns which map the checksum code, are connected to the same input as the one of the crossbars, they compute an additional MAC operation, whose output should correspond to the one previously calculated. If the computation is correct, the two values will match. Otherwise, a fault may have occurred in either the crossbar or the safety block. 
\subsection{PE Checksum}
\begin{figure}[htb]
\centering{\resizebox{\columnwidth}{!}{\input{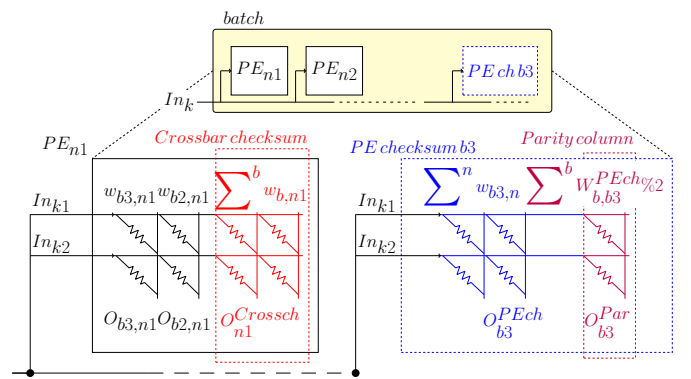}}} 
\caption{Architectural representation of the two checksums for a $n$-PEs batch.} \label{fig:myarch}
\end{figure}

This code is called "PE checksum" ($O_b^{PEch}$) as it is implemented in the form of redundant PE.
Each additional block maps the values that correspond to the sum of parameters belonging to the same column index of different crossbars within the same row of PEs ($W^{PEch}$), as illustrated in Fig. \ref{fig:myarch}. The same set of inputs is forwarded to each crossbar of the same row of PEs. This process is extended to the redundant PE that maps the checksum code computing the output in parallel. After the computation is complete, the results from the different crossbars having the same column index, are digitally added using adder trees, and compared with those of the checksum. If a mismatch is detected, it can be inferred that a fault occurred in one or multiple crossbar columns or the safety blocks.
\subsection{IMC Error Detection and Correction Routine}
\begin{figure}[htb]
\centering
\includegraphics[width= \columnwidth]{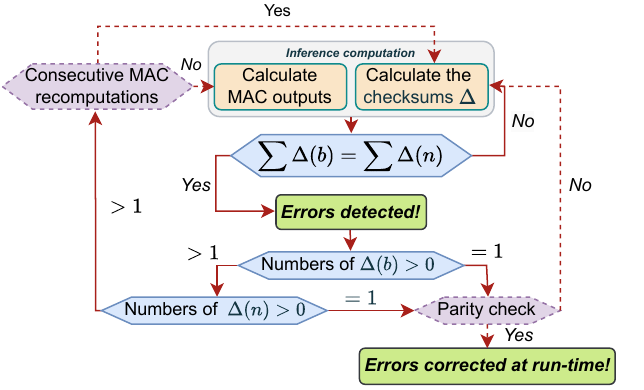}
\caption{Flowchart of the IEDCR.} \label{fig:err}
\end{figure}

In order to safely locate the part of the design where the fault occurred, the "IMC Error Detection and Correction Routine" (IEDCR) in Fig. \ref{fig:err} has been implemented.
Once a fault event is triggered as a consequence of a detected mismatch between the outputs and the redundant blocks, the differences between the two checksums and the original values are added together, and then the two resulting numbers are compared. In case they are equal then it is likely that the checksums computed correctly and a fault occurred somewhere within the device.  Otherwise, it is reasonable to assume that a malfunction occurred in the safety blocks, and therefore, a stall should be introduced to repeat their calculations. Once the correct computation of the checksum is verified, the indexes of the corresponding safety blocks are used as coordinates to determine in which crossbar column(s) of which PE(s) a fault occurred. The index $n$ of $O_n^{Crossch}$, for instance, identifies the faulty PE, while the index $b$ of $O_b^{PEch}$ corresponds to the specific faulty crossbar's column.
If only one crossbar column is faulty, the result can be corrected by adding the computed difference ($\Delta$) for the $O_n^{Crossch}$ to the output of the crossbar’s column of the corresponding PE(s). In this case, each PE's $\Delta$ corresponds to the error occurring in the specific column $b$ of the corresponding crossbar, as shown in the equations \ref*{eq:faultcol} and \ref*{eq:correctioncol}.
\begin{equation}
    \label{eq:faultcol}
    \resizebox{1.04\columnwidth}{!}{$
    \begin{cases}
        \Delta(b) = O_b^{PEch} - O_b^{acc}= \Delta(O_{b,n1},O_{b,n1}^{fault}) + \Delta(O_{b,n2},O_{b,n2}^{fault}) \\ 
        \Delta(n1) = O_{n1}^{Crossch} - O_{n1}^{acc}= \Delta(O_{b,n1},O_{b,n1}^{fault}) \\
        \Delta(n2) = O_{n2}^{Crossch} - O_{n2}= \Delta(O_{b,n2},O_{b,n2}^{fault})\\
    \end{cases}
    $}
\end{equation}
if ($\Delta(n1) + \Delta(n2) == \Delta(b)$) then:
\begin{align}
    \label{eq:correctioncol}
    \begin{split}
        &O_{b,n1}^{corrected} := O_{b,n1} +\Delta(n1)\\
        &O_{b,n2}^{corrected} := O_{b,n2} +\Delta(n2)
    \end{split}
\end{align}
If there are multiple faulty crossbar columns and only one faulty PE, the fault can still be corrected by adding the $\Delta$ computed for the $O_b^{PEch}$ of the corresponding $b$ column(s) to the outputs of the crossbar's column(s) within the single faulty PE. In this case, the $\Delta$ computed for each column corresponds to the error occurring within the specific $n$ PE, as shown in the equations \ref*{eq:faultpe} and \ref*{eq:correctionpe}.
\begin{equation}
    \label{eq:faultpe}
    \resizebox{1.02\columnwidth}{!}{$
    \begin{cases}
        \Delta(n) = O_{n}^{Crossch}- O_{n}^{acc}= \Delta(O_{b1,n},O_{b1,n}^{fault}) + \Delta(O_{b2,n},O_{b2,n}^{fault}) \\
        \Delta(b1) = O_{b1}^{PEch} - O_{b1}^{acc}= \Delta(O_{b1,n},O_{b1,n}^{fault}) \\
        \Delta(b2) = O_{b2}^{PEch} - O_{b2}^{acc}= \Delta(O_{b2,n},O_{b2,n}^{fault})\\
    \end{cases}
    $}
\end{equation}
if ($\Delta(b1) + \Delta(b2) == \Delta(n)$) then:
\begin{align}
    \label{eq:correctionpe}
    \begin{split}
        &O_{b1,n}^{corrected} := O_{b1,n} + \Delta(b1)\\
        &O_{b2,n}^{corrected} := O_{b2,n} + \Delta(b2)
    \end{split}
\end{align}
If both the number of faulty crossbar columns and PEs is greater than one, then the errors cannot be corrected at run-time, but other techniques should be implemented. 
In this work, we opted to repeat the computation in the original device up to the minimum number of cycles required to obtain one of the previously described scenarios. This provides the hardware designers with an additional parameter when evaluating the overhead for the configuration best suited to their use case.
\subsection{Malfunctions in the Checksum Blocks}
If a fault distorts the calculations of both safety blocks with the same arithmetic error, it may result in either an erroneous correction or an endless introduction of stalls in the design. To avoid these scenarios, an additional stall is issued to recalculate the checksums' values either after a fixed number of consecutive stalls or when the output of the column mapping the parity codes of the "PE checksum" crossbar is inconsistent.
\section{Experimental Setup and Results}
\label{sec:results}
\begin{figure*}[htb]
\centering 
\begin{subfigure}{0.5\textwidth}
  \includegraphics[width=\linewidth]{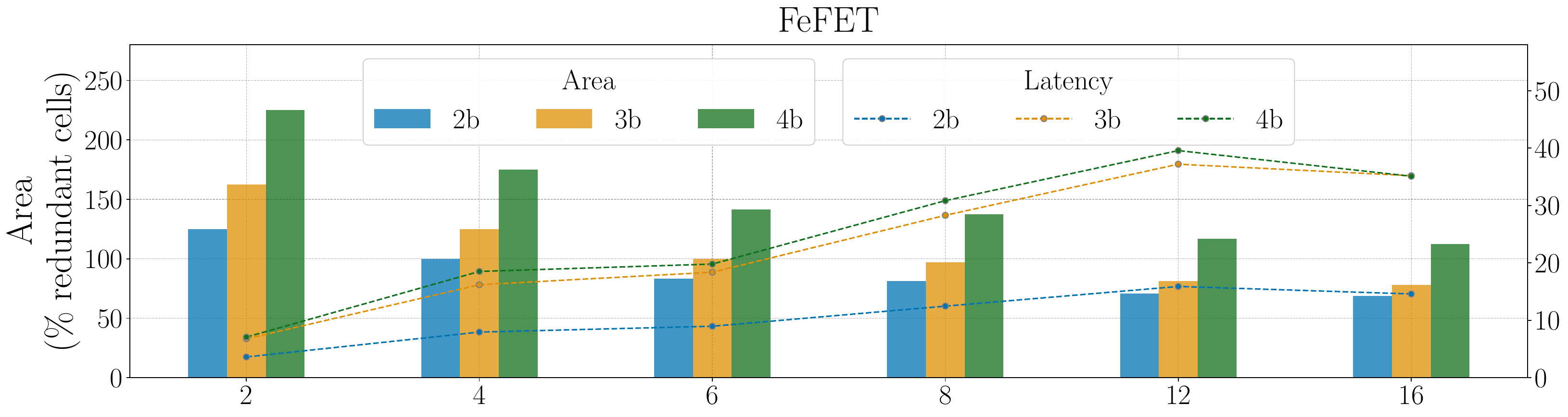}
\end{subfigure}\hfil
\begin{subfigure}{0.5\textwidth}
  \includegraphics[width=\linewidth]{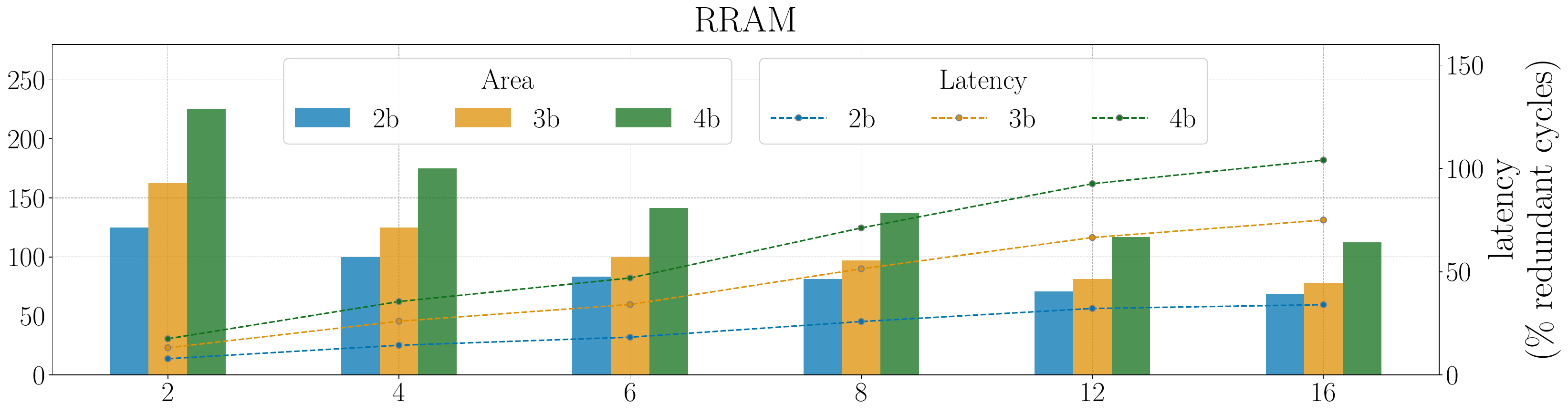}
\end{subfigure}\hfil
\caption{Checksum overhead in terms of area and latency for the Resnet20 NN architecture (left: FeFET, right: RRAM).}
\vspace{-2 pt}
\label{fig:overhead}
\end{figure*}

Several experiments were conducted using ProxSim \cite{de_la_parra_proxsim_2020}, a GPU-based framework for hardware-aware retraining and evaluation of DNN computations. The error model utilized in this study simulates the effects of soft faults within an IMC architecture. It was implemented by injecting stochastic errors at runtime to generate arithmetic deviations in the output of each column of a crossbar.
We ran the inference phase for the Resnet20 \cite{he2015deep}, Resnet32 \cite{he2015deep}, and Neural in Network (NiN) \cite{lin2014network} algorithms trained on the CIFAR10 dataset \cite{krizhevsky_convolutional_nodate}. In order to show the applicability of our safety mechanism to highly compressed networks the quantizations chosen for the weights and input activations were 4-bit and 8-bit signed integers, respectively.
As shown in Fig. \ref{fig:overhead}, for the Resnet20 algorithm, we conducted additional tests to reduce the overhead needed by the safety blocks while maintaining a satisfactory level of accuracy. In these experiments, we focused on protecting specific logic sections that store the most critical parameters for the computation of accurate predictions.
More specifically, we adopted a blind approach to rank the criticality of the logic according to the significance of the weights' bits it represents, starting from two, three, and finally a complete representation of four bits. 
To further investigate the tradeoffs that can be achieved in terms of area and latency overhead, we ran the experiments for multiple configurations of the checksums.
To illustrate, we can divide the matrix of PEs within the device into separate batches. Each batch implements the corresponding checksums and independently carries out its own IEDCR. By increasing the number of PEs within each batch, the overall overhead in terms of the area required by the checksums decreases, while the latency increases due to the insertion of additional correction cycles. This is due to the higher probability of multiple errors in the protected logic.

\begin{table*}[ht]
\centering
\caption{Normalized accuracy, error detection and correction rates for different technologies and bit protections.}
\arrayrulecolor[rgb]{0.753,0.753,0.753}
\resizebox{0.9\textwidth}{!}{
\begin{tabular}{!{\color{black}\vrule}c!{\color{black}\vrule}c|c|c|c|c|c!{\color{black}\vrule}c|c|c|c|c|c!{\color{black}\vrule}} 
\arrayrulecolor{black}\hline
{\cellcolor[rgb]{0.859,0.859,0.671}}                                                                                                                                                                & \multicolumn{6}{c!{\color{black}\vrule}}{{\cellcolor[rgb]{0.902,0.929,1}}\textbf{FeFET 500x500}}                                                                                                                                                                                                                                                                                                                                                                                                                                                                                                                                                                                                                                                                                        & \multicolumn{6}{c!{\color{black}\vrule}}{{\cellcolor[rgb]{0.71,0.851,0.71}}\textbf{RRAM}}                                                                                                                                                                                                                                                                                                                                                                                                                                                                                                                                                                                                                                                                                                                                                                                                                                                                                                                        \\ 
\hhline{|>{\arrayrulecolor[rgb]{0.859,0.859,0.671}}->{\arrayrulecolor{black}}------------|}
\rowcolor[rgb]{0.902,0.929,1} \multirow{-2}{*}{{\cellcolor[rgb]{0.859,0.859,0.671}}\begin{tabular}[c]{@{}>{\cellcolor[rgb]{0.859,0.859,0.671}}c@{}}\textbf{Neural }\\\textbf{Network}\end{tabular}} & \multicolumn{1}{c!{\color{black}\vrule}}{\begin{tabular}[c]{@{}>{\cellcolor[rgb]{0.902,0.929,1}}c@{}}Original\\accuracy\end{tabular}} & \multicolumn{1}{c!{\color{black}\vrule}}{\begin{tabular}[c]{@{}>{\cellcolor[rgb]{0.902,0.929,1}}c@{}}TMR\\accuracy\end{tabular}} & \multicolumn{1}{c!{\color{black}\vrule}}{\begin{tabular}[c]{@{}>{\cellcolor[rgb]{0.902,0.929,1}}c@{}}Bits\\protected\end{tabular}} & \multicolumn{1}{c!{\color{black}\vrule}}{\begin{tabular}[c]{@{}>{\cellcolor[rgb]{0.902,0.929,1}}c@{}}Checksums\\accuracy\end{tabular}} & \multicolumn{1}{c!{\color{black}\vrule}}{\begin{tabular}[c]{@{}>{\cellcolor[rgb]{0.902,0.929,1}}c@{}}Error\\detected\end{tabular}} & \begin{tabular}[c]{@{}>{\cellcolor[rgb]{0.902,0.929,1}}c@{}}Error\\corrected\end{tabular} & \multicolumn{1}{c!{\color{black}\vrule}}{{\cellcolor[rgb]{0.71,0.851,0.71}}\begin{tabular}[c]{@{}>{\cellcolor[rgb]{0.71,0.851,0.71}}c@{}}Original\\accuracy\end{tabular}} & \multicolumn{1}{c!{\color{black}\vrule}}{{\cellcolor[rgb]{0.71,0.851,0.71}}\begin{tabular}[c]{@{}>{\cellcolor[rgb]{0.71,0.851,0.71}}c@{}}TMR\\accuracy\end{tabular}} & \multicolumn{1}{c!{\color{black}\vrule}}{{\cellcolor[rgb]{0.71,0.851,0.71}}\begin{tabular}[c]{@{}>{\cellcolor[rgb]{0.71,0.851,0.71}}c@{}}Bits\\protected\end{tabular}} & \multicolumn{1}{c!{\color{black}\vrule}}{{\cellcolor[rgb]{0.71,0.851,0.71}}\begin{tabular}[c]{@{}>{\cellcolor[rgb]{0.71,0.851,0.71}}c@{}}Checksums\\accuracy\end{tabular}} & \multicolumn{1}{c!{\color{black}\vrule}}{{\cellcolor[rgb]{0.71,0.851,0.71}}\begin{tabular}[c]{@{}>{\cellcolor[rgb]{0.71,0.851,0.71}}c@{}}Error\\detected\end{tabular}} & {\cellcolor[rgb]{0.71,0.851,0.71}}\begin{tabular}[c]{@{}>{\cellcolor[rgb]{0.71,0.851,0.71}}c@{}}Error\\corrected\end{tabular}  \\ 
\hline
\rowcolor[rgb]{0.973,0.976,1} {\cellcolor[rgb]{1,1,0.78}}                                                                                                                                           & {\cellcolor[rgb]{0.973,0.976,1}}                                                                                                      & {\cellcolor[rgb]{0.973,0.976,1}}                                                                                                 & 2                                                                                                                                  & 67.60                                                                                                                                  & 51.31                                                                                                                              & 51.30                                                                                     & {\cellcolor[rgb]{0.925,0.961,0.925}}                                                                                                                                      & {\cellcolor[rgb]{0.925,0.961,0.925}}                                                                                                                                 & {\cellcolor[rgb]{0.925,0.961,0.925}}2                                                                                                                                  & {\cellcolor[rgb]{0.925,0.961,0.925}}73.68                                                                                                                                  & {\cellcolor[rgb]{0.925,0.961,0.925}}47.24                                                                                                                              & {\cellcolor[rgb]{0.925,0.961,0.925}}44.05                                                                                      \\ 
\hhline{|>{\arrayrulecolor[rgb]{1,1,0.78}}->{\arrayrulecolor[rgb]{0.973,0.976,1}}-->{\arrayrulecolor[rgb]{0.753,0.753,0.753}}---->{\arrayrulecolor[rgb]{0.925,0.961,0.925}}-->{\arrayrulecolor[rgb]{0.753,0.753,0.753}}---->{\arrayrulecolor{black}}|}
\rowcolor[rgb]{0.973,0.976,1} {\cellcolor[rgb]{1,1,0.78}}                                                                                                                                           & {\cellcolor[rgb]{0.973,0.976,1}}                                                                                                      & {\cellcolor[rgb]{0.973,0.976,1}}                                                                                                 & 3                                                                                                                                  & 91.55                                                                                                                                  & 75.19                                                                                                                              & 75.18                                                                                     & {\cellcolor[rgb]{0.925,0.961,0.925}}                                                                                                                                      & {\cellcolor[rgb]{0.925,0.961,0.925}}                                                                                                                                 & {\cellcolor[rgb]{0.925,0.961,0.925}}3                                                                                                                                  & {\cellcolor[rgb]{0.925,0.961,0.925}}91.34                                                                                                                                  & {\cellcolor[rgb]{0.925,0.961,0.925}}69.60                                                                                                                              & {\cellcolor[rgb]{0.925,0.961,0.925}}69.52                                                                                      \\ 
\hhline{|>{\arrayrulecolor[rgb]{1,1,0.78}}->{\arrayrulecolor[rgb]{0.973,0.976,1}}-->{\arrayrulecolor[rgb]{0.753,0.753,0.753}}---->{\arrayrulecolor[rgb]{0.925,0.961,0.925}}-->{\arrayrulecolor[rgb]{0.753,0.753,0.753}}---->{\arrayrulecolor{black}}|}
\rowcolor[rgb]{0.973,0.976,1} \multirow{-3}{*}{{\cellcolor[rgb]{1,1,0.78}}\begin{tabular}[c]{@{}>{\cellcolor[rgb]{1,1,0.78}}c@{}}\textit{Resnet20}\end{tabular}}               & \multirow{-3}{*}{{\cellcolor[rgb]{0.973,0.976,1}}12.48}                                                                               & \multirow{-3}{*}{{\cellcolor[rgb]{0.973,0.976,1}}15.24}                                                                          & 4                                                                                                                                  & 98.75                                                                                                                                  & 99.94                                                                                                                              & 99.93                                                                                     & \multirow{-3}{*}{{\cellcolor[rgb]{0.925,0.961,0.925}}64.09}                                                                                                               & \multirow{-3}{*}{{\cellcolor[rgb]{0.925,0.961,0.925}}84.82}                                                                                                          & {\cellcolor[rgb]{0.925,0.961,0.925}}4                                                                                                                                  & {\cellcolor[rgb]{0.925,0.961,0.925}}95.81                                                                                                                                  & {\cellcolor[rgb]{0.925,0.961,0.925}}96.74                                                                                                                              & {\cellcolor[rgb]{0.925,0.961,0.925}}96.66                                                                                      \\ 
\hline
\rowcolor[rgb]{0.973,0.976,1} {\cellcolor[rgb]{1,1,0.78}}\begin{tabular}[c]{@{}>{\cellcolor[rgb]{1,1,0.78}}c@{}}\textit{NiN}\end{tabular}                                     & 15.45                                                                                                                                 & 22.78                                                                                                                            & 4                                                                                                                                  & 99.96                                                                                                                                  & 99.96                                                                                                                              & 99.94                                                                                     & {\cellcolor[rgb]{0.925,0.961,0.925}}74.84                                                                                                                                 & {\cellcolor[rgb]{0.925,0.961,0.925}}91.41                                                                                                                            & {\cellcolor[rgb]{0.925,0.961,0.925}}4                                                                                                                                  & {\cellcolor[rgb]{0.925,0.961,0.925}}98.37                                                                                                                                  & {\cellcolor[rgb]{0.925,0.961,0.925}}96.39                                                                                                                              & {\cellcolor[rgb]{0.925,0.961,0.925}}96.32                                                                                      \\ 
\hline
\rowcolor[rgb]{0.973,0.976,1} {\cellcolor[rgb]{1,1,0.78}}\begin{tabular}[c]{@{}>{\cellcolor[rgb]{1,1,0.78}}c@{}}\textit{Resnet32 }\end{tabular}                                & 10.83                                                                                                                                 & 12.43                                                                                                                            & 4                                                                                                                                  & 99.56                                                                                                                                  & 99.96                                                                                                                              & 99.96                                                                                     & {\cellcolor[rgb]{0.925,0.961,0.925}}85.24                                                                                                                                 & {\cellcolor[rgb]{0.925,0.961,0.925}}96.66                                                                                                                            & {\cellcolor[rgb]{0.925,0.961,0.925}}4                                                                                                                                  & {\cellcolor[rgb]{0.925,0.961,0.925}}98.82                                                                                                                                  & {\cellcolor[rgb]{0.925,0.961,0.925}}96.72                                                                                                                              & {\cellcolor[rgb]{0.925,0.961,0.925}}96.66                                                                                      \\
\hline
\end{tabular}
}
\vspace{-6 pt}
\label{tab:my-table}
\end{table*}
\subsection{Results and Overhead}
Table \ref{tab:my-table} presents the accuracy percentage and error detection and correction rates for various experiments, with and without protection mechanisms,
including TMR, for reference. Fig. \ref{fig:overhead} displays the area and latency overheads, revealing a pattern that hardware designers can utilize to choose the best configuration that meets the constraints for their specific use case. According to the results, TMR demonstrates a better error correction performance in the context of RRAM technologies.
On the other hand, the accuracy achieved with our checksums remains high in both cases, allowing for retrieval of up to 95\% of the original value. This behavior confirms our approach's greater flexibility and technology agnosticism.
In terms of the 200\% increase in redundant cells necessary for TMR, our design is optimized for the majority of $n$-PEs batches and bits protection. 
Table \ref{tab:my-table_ov} also shows that our designs require an area overhead comparable to the solutions proposed by the state-of-the-art. The values presented in Table \ref{tab:my-table_ov} have been adjusted to our context, where each cell represents a binary state instead of multiple bits. Comparing the error correction capability or accuracy retrieval is more challenging, as the error models used to test different designs differ from ours. For instance \cite{liu_fault_2018}, reports the performance of their design for different scenarios with increasing probabilities of having a random fault. For 5\% of faulty RRAM cells, they report up to 67\% of corrections, while in our case, we can reach up to 96\%.
The latency requirements are highly dependent on the configuration and technology being tested.
For example, using the configuration of 12 PEs per batch and three bits protection, we achieve comparable, if not superior, results in terms of accuracy while incurring less than half of the area overhead necessary for TMR and less than 40\% and 75\% latency overhead for FeFET and RRAM technologies, respectively.
\begin{table}[h]
\centering
\vspace{10 pt}
\caption{Percentage redundant cells for different designs.}
\begin{tabular}{c|l|l|lll}
\hline
\multirow{2}{*}{\cite{liu_fault_2018}}   & \multirow{2}{*}{\hspace{0.25 cm}\cite{hemaram_adaptive_2022}}   & \multirow{2}{*}{\hspace{0.25 cm}\cite{das_selective_2020}}   & \multicolumn{3}{c}{proposed design} \\ \cline{4-6} 
                            &                             &                             & \multicolumn{1}{l|}{2 bits}     & \multicolumn{1}{l|}{3 bits}    & 4 bits     \\ \hline
\multicolumn{1}{l|}{200-31} & \multicolumn{1}{l|}{192-55} & \multicolumn{1}{l|}{131-56} & \multicolumn{1}{l|}{125-69}     & \multicolumn{1}{l|}{162-78}    & 225-112    \\ \hline
\end{tabular}
\label{tab:my-table_ov}

\end{table}

\section{Conclusion}
\label{sec:conclusions}
In this study, we presented a mitigation technique for detecting and correcting errors through the implementation of checksum codes in IMC architectures with RRAM and FeFET technologies. Multiple tests were conducted for three algorithms when safety measures were not implemented, as well as TMR with a voting system and various checksum configurations.
The results show that our design maintains superior accuracy with higher fault detection and correction rates in both technologies at the cost of half the area overhead of traditional methods such as TMR.
\section*{Acknowledgment}
This work has been funded by the Federal Ministry of Education and Research (BMBF) project MANNHEIM-FlexKI (01IS22086A).
\vspace{-2 pt}

\bibliographystyle{IEEEtran}

\bibliography{bstcontrol, myLibrary}    

\end{document}